\documentclass[10pt,twocolumn,english,preprint,prb,tightenlines]{revtex4}
\usepackage[T1]{fontenc}
\usepackage[latin9]{inputenc}
\usepackage{textcomp}
\usepackage{amsmath}
\usepackage{amssymb}
\usepackage{graphicx}

\makeatletter
\@ifundefined{textcolor}{}
{%
 \definecolor{BLACK}{gray}{0}
 \definecolor{WHITE}{gray}{1}
 \definecolor{RED}{rgb}{1,0,0}
 \definecolor{GREEN}{rgb}{0,1,0}
 \definecolor{BLUE}{rgb}{0,0,1}
 \definecolor{CYAN}{cmyk}{1,0,0,0}
 \definecolor{MAGENTA}{cmyk}{0,1,0,0}
 \definecolor{YELLOW}{cmyk}{0,0,1,0}
 }

\makeatother

\usepackage{babel}
\begin{document}

\title{Electronically driven superconductor insulator transition in electrostatically
doped La$_{2}$CuO$_{4+\delta}$ thin films}

\author{{\normalsize J. Garcia-Barriocanal$^{1,2}$, A. Kobrinskii$^{1}$,
X. Leng$^{1}$, J. Kinney$^{1}$, B. Yang$^{1}$, S. Snyder$^{1}$
and A. M Goldman$^{1}$}}

\address{$^{1}$School of Physics and Astronomy, University of Minnesota,
Minneapolis, Minnesota 55455, USA}

\address{$^{2}$GFMC, Dpto. de Fisica Aplicada III, Universidad Complutense
de Madrid, 28040 Madrid, Spain}
\begin{abstract}
Using an electronic double layer transistor we have systematically
studied the superconductor to insulator transition in La$_{2}$CuO$_{4+\delta}$
thin films growth by ozone assisted molecular beam epitaxy. We have
confirmed the high crystalline quality of the cuprate films and have
demonstrated the suitability of the electronic double layer technique
to continuously vary the charge density in a system that is otherwise
characterized by the presence of miscibility gaps. The transport and
magneto-transport results highlight the role of electron-electron
interactions in the mechanism of the transition due to the proximity
of the Mott insulating state.
\end{abstract}
\maketitle

\section{Introduction}

Despite extensive efforts, a generally accepted theory of the superconductivity
of cuprate superconductors is still lacking.\cite{Leggett2006,Lee2006}
The description of the superconductor to insulator transition (SIT)
has not been completed for these materials, leaving important questions
concerning the origin of the High Temperature Superconducting (HTS)
state, and the phases that are found in its intriguing phase diagram,
unresolved.\cite{Varma2010} The superconducting condensate in cuprates
emerges below a critical temperature (T$_{C}$) with the addition
of charge carriers to an antiferromagnetic Mott Insulator (MI). It
is generally accepted that in the vicinity of the Mott state the essential
2D character of the superconducting CuO$_{2}$ planes gives rise to
new properties that cannot be explained by Fermi liquid theory.\cite{Orenstein2000}
Interestingly, this unconventional behavior is not limited to the
superconducting low temperature phase of the cuprates. It also affects
the \textquotedblleft{}normal\textquotedblright{} high temperature
regime denoted as the pseudogap (PG). Understanding the mechanisms
that turn MIs into HTSs, specifically in the PG region, remains one
of the great intellectual challenges of condensed matter physics.

The SIT is one example of a quantum phase transition (QPT), which
occurs at zero temperature.\cite{Sondhi1997} In disordered ultrathin
metal films it has been interpreted with either a bosonic or a femionic
description and the consequences of each picture are different. On
the one hand, the bosonic picture involves the presence of Cooper
pairs (i.e. the charge carriers of the superconducting state) and
vortices (their resistive counterparts) in both the insulating and
superconducting phases.\cite{Fisher1990} The Cooper pairs are localized
and the vortices are delocalized in the insulating regime, and the
Cooper pairs condense into a charged superfluid on the superconducting
side of the transition while vortices become localized. On the other
hand, the fermionic picture implies the destruction of superconductivity
at the SIT by means of pair breaking. Accordingly, it can be expected
that the charge carriers on the insulating side of the transition
are single electrons or holes, and thus, the fermionic picture would
highlight the importance of electronic correlations possibly due to
the proximity of the Mott insulator.\cite{Anderson1986}

Here we report the systematic study of the SIT of a four unit cell
thick ultrathin film of La$_{2}$CuO$_{4+\delta}$ ($\delta$-LCO)
using electrostatically induced charge as the tuning parameter of
the transition. This has been facilitated by the combination of the
Ozone assisted Molecular Beam Epitaxy (OMBE) growth technique together
with the fabrication of an Electronic Double Layer (EDL) device. OMBE
provides high crystalline quality and flat ultrathin films of a robust
$\delta$-LCO superconductor (T$_{C}$ $\sim$ 45 K as measured at
the onset of the transition) grown under compressive epitaxial strain
on top of SrLaAlO$_{4}$ (SLAO) substrates. The EDL technique employing
ionic liquids (ILs) has been used to successfully induce levels of
doping of the order of 10$^{15}$ cm$^{-2}$ ,\cite{Shimotani2007,Ye2010,Lee2011}
and it allows for the systematic and exhaustive study of the SIT.
This approach avoids crystal inhomogeneity due to phase separation
and does not change the random Coulomb potential at the different
levels of doping. In this regard, the work presented here would address
some of the fundamental issues of HTS. Among other results, in-plane
low temperature anisotropy of the electronic properties of the film
has been found. This anisotropy is a minimum at the SIT, which occurs
at the same hole concentration as a maximum in the Hall resistance
measured at high temperatures, suggesting that the SIT of electrostatically
doped $\delta$-LCO films is electronically driven. However, it should
be noted that the system under study differs from the bulk due to
its low dimensionality, the presence of high electric fields at its
surface, and the very different method of doping. As such it could
become a source of interesting new physical phenomena.

\section{Delta Doped Lanthanum Cuprate ($\delta$-LCO)}

Stoichiometric La$_{2}$CuO$_{4}$ (LCO) is an antiferromagnetic Mott
insulator and is one of the most extensively studied precursors of
HTS (as well as the HTS families derived from it) .\cite{Kastner1998,Imada1998,Lee2006}
Among those families of doped LCO cuprates, the oxygen doped compounds
(La$_{2}$CuO$_{4+\delta}$ {[}$\delta$-LCO{]}) stand out for their
annealed disorder vs. the quench disorder that arises in cation substituted
compounds. The interstitial oxygens (i-O) in $\delta$-LCO compounds
are located in the spacer La$_{2}$O$_{2+\delta}$ layers that are
intercalated between the CuO$_{2}$ layers. These added oxygens are
mobile down to relatively low temperatures ( > 200 K) and they are
able to find an equilibrium arrangement in the crystallographic structure.\cite{Wells1997,Lee1999,Fratini2010}
Interstitial oxygens are located at the (\textonequarter{}, \textonequarter{},
\textonequarter{}) crystallographic positions and they can form an
\textquotedblleft{}ordered sublattice\textquotedblright{} under appropriate
conditions of synthesis and treatment of the crystal. They exhibit
staging order in the direction perpendicular to the CuO$_{2}$ planes,
and they form in-plane stripes along the main diagonal of the tetragonal
unit cell breaking the symmetry of the Cu-O octahedra along the main
in-plane directions of the pseudocubic unit cell.\cite{Chaillout1989}
The expected pattern of disorder in this family of compounds is therefore
weaker than that obtained for chemically substituted compounds.\cite{Eisaki2004}
$\delta$-LCO crystals are believed to be the simplest cuprate HTS
since the i-O are not randomly distributed, and moreover, they have
the highest T$_{C}$ since the i-O are located far away from the superconducting
CuO$_{2}$ planes.\cite{Wells1997,Eisaki2004} Unfortunately, in the
study of the phase diagram of $\delta$-LCO bulk compounds it is impossible
to produce a series of doped samples in which the concentration of
oxygen is systematically varied since the system exhibits phase separation
over wide ranges of oxygen doping.\cite{Wells1997}

\section{Epitaxial growth and structural characterization of $\delta$-LCO
thin films}

\subsection{Ozone assisted molecular beam epitaxy}

Epitaxial $\delta$-LCO thin films have been grown by means of the
OMBE technique. Pure ozone is distilled from a mixture of O$_{2}$/O$_{3}$
gas which is produced with a silent discharge ozone generator. The
ozone gas is supplied to the chamber at a constant pressure value
of 3 x 10$^{-5}$ Torr during growth. Further details about the MBE
system and the sample preparation can be found in previous work.\cite{Berkley1989}
The samples of the present study were grown on (001) oriented SrLaAlO$_{4}$
(SLAO) substrates at 750 ºC with a shuttered growth technique. Samples
were slowly cooled, over 45 minutes, down to 50 ºC in an ozone atmosphere,
to prevent de-oxygenation. In order to characterize the growth mechanism,
test samples were also grown on (001) oriented SrTiO$_{3}$ (STO)
substrates. The results from the \textit{in situ} control of the growth
and the structural characterization experiments were the same in terms
of crystal quality, for films grown on top of both SLAO and STO substrates.

\begin{figure}
\includegraphics[width=7.5cm]{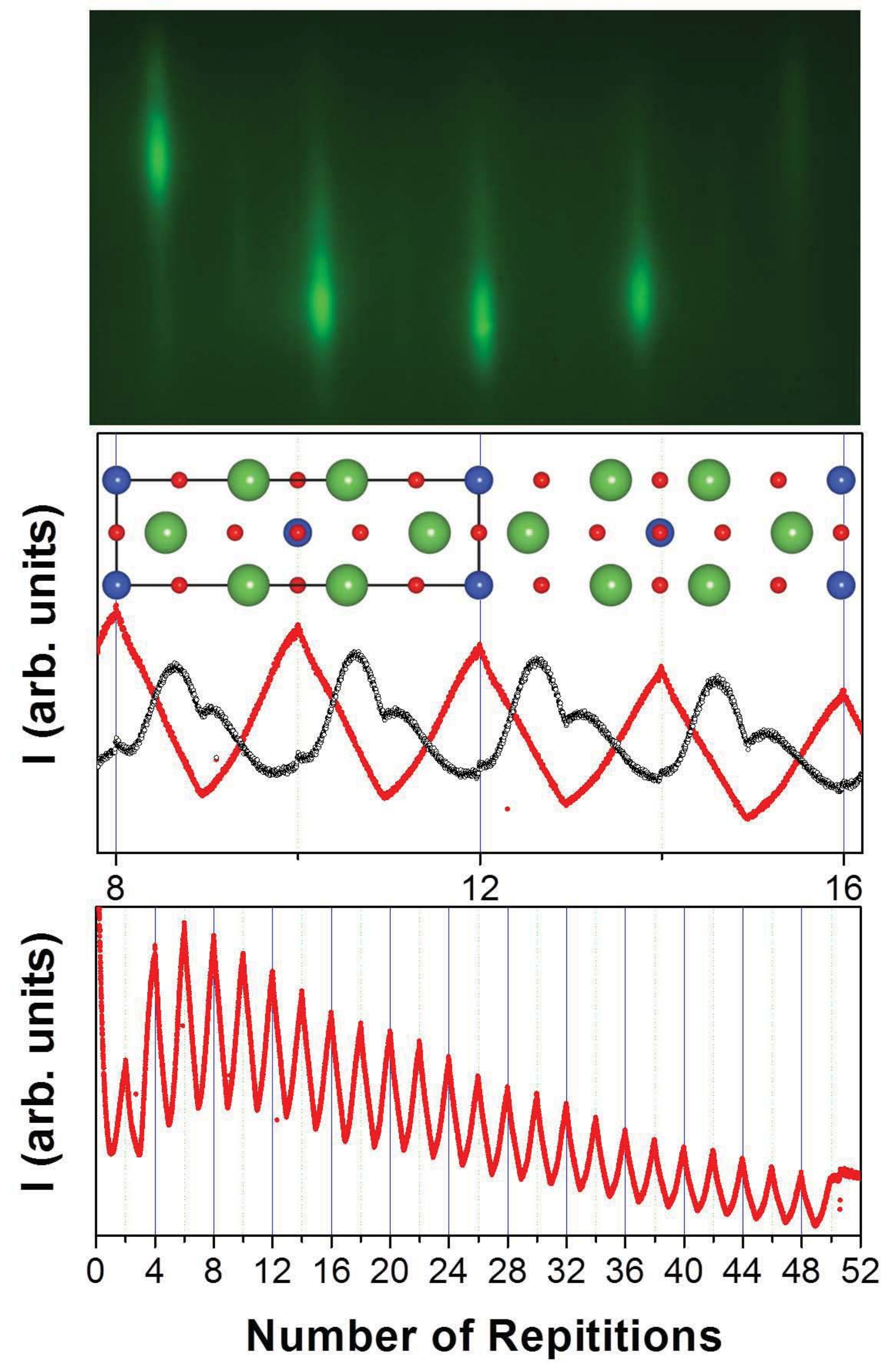}\caption{(Color online) The top image is an example of the RHEED pattern obtained
after the growth of a 4 unit cell thick LCO thin film. The bottom
graph shows the oscillation of the integrated intensity of the specular
rod (red line) during the growth of a 12.5 unit cell sample as a function
of the number of repetitions (odd numbers correspond to the opening
of the Cu shutter and even numbers to the La shutter). The top graph
shows the details of the 4th and 5th unit cells of the same growth
and it represents the single atomic plane control of the growth (see
text). The $\delta$-LCO crystallographic structure is sketched on
top of this panel to better visualize the growth sequence.}
\end{figure}

The growth sequence starts with the deposition of a La layer followed
by a Cu layer. \textit{In situ} control of the layer-by-layer growth
is provided by Reflection High Energy Electron Diffraction (RHEED)
patterns acquired during the growth process (top image of Fig. 1).
After the deposition of the La layer a three-dimensional (3D) RHEED
pattern, characteristic of a rough surface, can be observed. Once
the deposition of the Cu layer is finished, the characteristic scattering
rods of a two-dimensional (2D) growth are revealed indicating the
presence of a flat surface (see top image of Fig. 1). In the bottom
graph of Fig. 1 we show the RHEED oscillations obtained from the integrated
intensity over the area of the specular rod during the growth of a
12.5 unit cell thick sample. Note that the time units of the X axis
of this graph have been scaled to the number of times that the La
and Cu shutters were opened. The count starts at zero which corresponds
to the first La deposition, and so, the odd (even) numbers in the
X axis of Fig. 1 correspond to the closing (opening) of the La shutter
and the opening (closing) of the Cu shutter. The RHEED oscillations
reveal that the growth method succeeds in preserving a smooth sample
surface up to the thickest ($\sim$300 A) films that were grown. Moreover,
one can follow the growth by counting the number of single atomic
planes of each element deposited. The middle graph of Fig. 1 shows
a detail of the same growth corresponding to the deposition of the
4th and 5th unit cells of the film. The black line in the figure is
obtained by integrating the intensity of the small spot that is characteristic
of the 3D growth of the RHEED pattern during the La and Cu deposition.
It reveals the formation of two LaO planes that takes place before
the completion of half of a single unit cell. The picture shown on
top of this graph is a guide to following the growth sequence and
it is a sketch of the crystallographic structure of LCO.

\subsection{Structural characterization of the LCO thin films}

The structural characterization of the films has been carried out
by means of X-Ray Diffraction and Atomic Force Microscopy (AFM). In
Fig. 2 we show the X-Ray results obtained for a set of characteristic
samples. Figure 2A shows the X-Ray Reflectivity (XRR) patterns of
12.5, 12, 4 and 4 unit cell thick LCO films grown on STO, SLAO, STO
and SLAO substrates respectively, from top to bottom. Finite size
thickness oscillations corresponding to the total thickness of the
film can be clearly observed over a wide range of values of 2$\theta$.
The presence of interference fringes indicates that both the substrate-sample
and sample-air interfaces are flat and smooth. A quantitative analysis
of the XRR results was carried out using GenX software.\cite{Bjorck2007}
The refinement results, the red lines in Fig. 2A are in agreement
with the nominal thicknesses observed during the growth using the
RHEED technique. We obtain a roughness of the sample-substrate interface
of less than 2 A, as well as a sample-air interface roughness of about
6 A which are in good agreement with the height of half a unit cell.
Both values of roughness are the same for all of the samples.

\begin{figure}
\includegraphics[width=8.6cm]{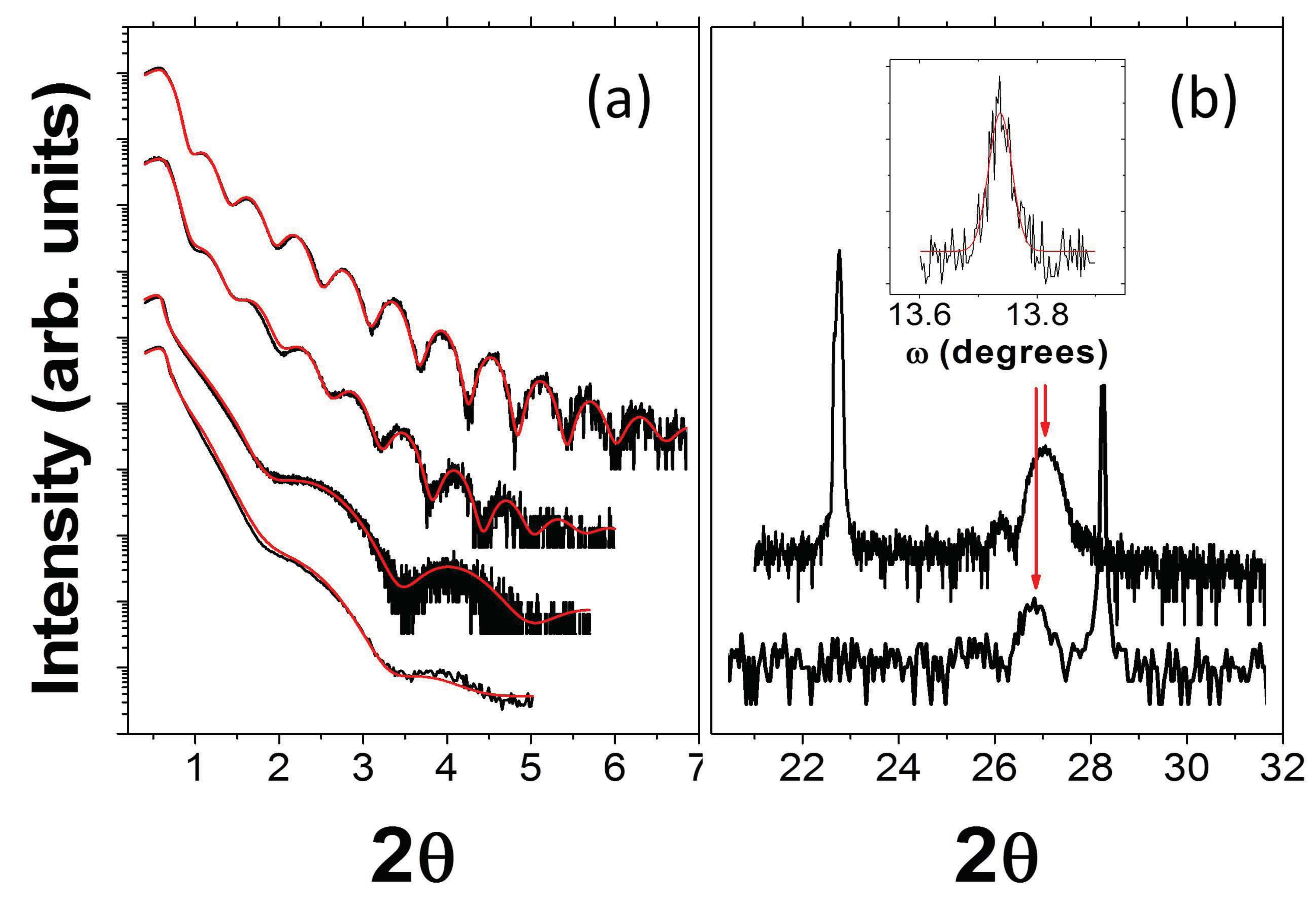}\caption{(Color online) (a) XRR data of four different $\delta$-LCO thin films,
from top to bottom: 12.5 unit cells grown on STO, 12 unit cells grown
on SLAO, 4 unit cells grown on STO and 4 unit cells grown on SLAO.
Both STO and SLAO are (0 0 1) oriented substrates. The red lines are
the refinement of the XRR data, see the text for the results of the
refinement. (b) Wide Angle X Ray Diffraction of the 12 and 12.5 unit
cells samples. In the inset the rocking curve of the first Bragg peak
of the film (red arrows in figure b) of the 12 unit cell sample is
shown as a representative example of all the samples of this work.
The rocking curve has a full width at half maximum of 0.03 degrees,
corresponding to a low crystallographic mosaicity with large domains.}
\end{figure}

The Wide Angle X-Ray Diffraction data (Fig. 2B) show that the sample
is grown textured in the (0 0 1) crystallographic direction of the
substrate with high crystalline quality. (From top to bottom we show
the data for the 12.5 and 12 unit cell samples grown on STO and SLAO
respectively.) No peaks of secondary phases can be seen over the whole
scan (12 < 2 $\theta$ < 65) but peaks corresponding to the (0 0 n)
family of crystallographic planes of both the substrates and $\delta$-LCO
thin films are observed. As marked with red arrows in Fig. 2B, the
position of the (0 0 4) Bragg peak of the thin film depends strongly
on the substrate used for the growth. As expected for growth on a
SLAO substrate, an in-plane compressively strained thin film with
a \textit{c-}axis lattice parameter c = 13.3 Å is obtained. 

\begin{figure}
\includegraphics[width=8.6cm]{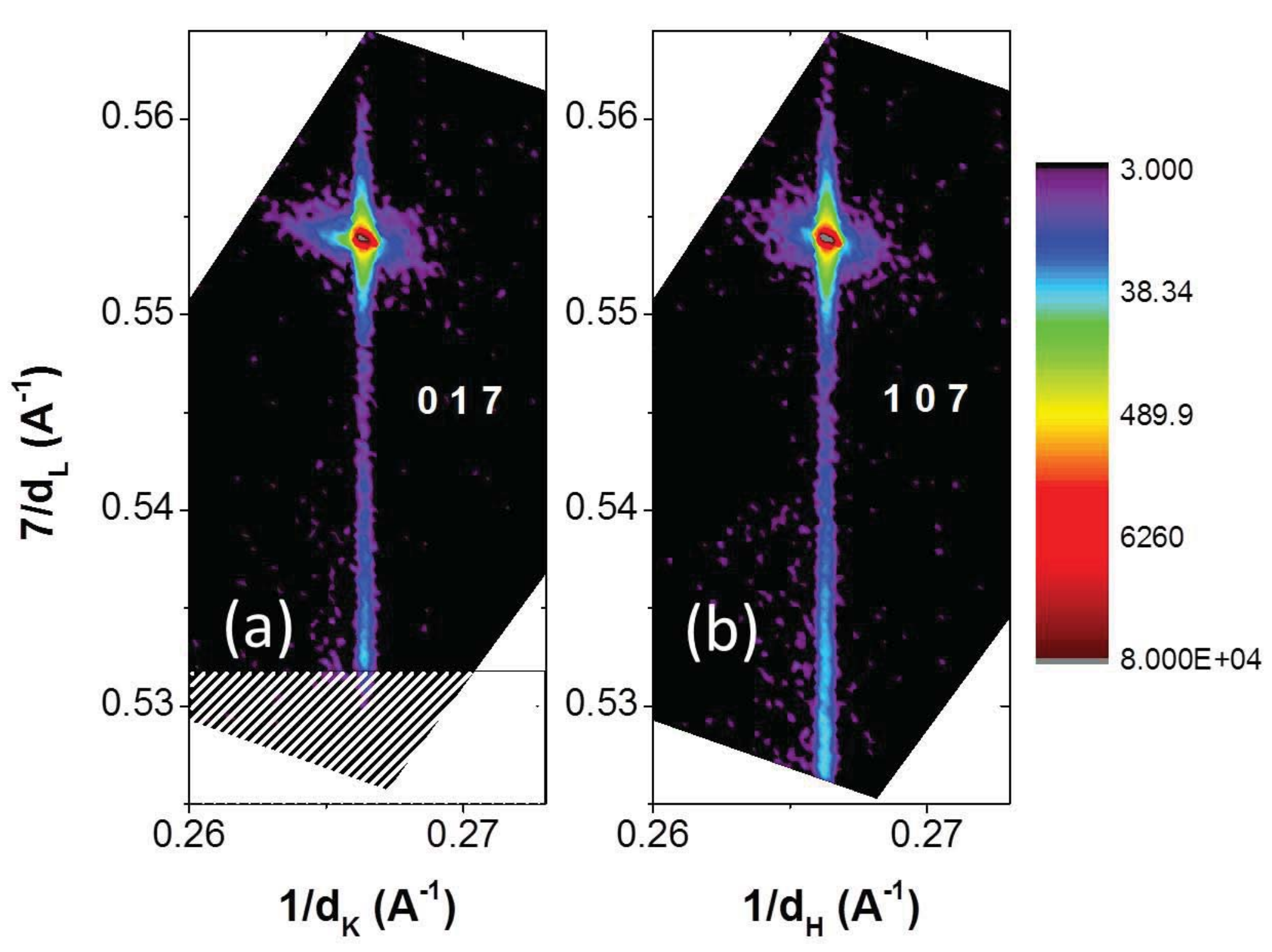}\caption{(Color online) Reciprocal space maps of the 4 unit cell $\delta$-LCO
thin film along the (0 1 7) and (1 0 7) crystallographic directions.}
\end{figure}

The presence of ozone and the compressive strain result in the formation
of a $\delta$-LCO thin film where the unit cell volume is not preserved
and a non-stoichiometric superconducting phase with excess of oxygen
is stabilized instead of the insulating antiferromagnetic stoichiometric
LCO phase. The introduction of oxygen interstitials at the La$_{2}$O$_{2+\delta}$
spacer layers together with the compressive strain produces an expansion
of the unit cell in the out-of-plane direction and turns the sample
into a high T$_{C}$ superconductor. The samples grown on STO are
tensile strained and have a c-axis parameter of 13.2 Å. With the samples
grown on STO we checked that the oxidizing atmosphere by itself was
able to stabilize $\delta$-LCO regardless of the strain state induced
in the thin film. In the inset of Fig. 2B we show the rocking curve
corresponding to the (0 0 4) Bragg peak of the 12 unit cell thick
film as a representative example. The full width at half maximum (FWHM)
of the peak is 0.03 degrees which indicates the very low mosaicity
of our samples and the structural coherence of the thin film in the
direction of the growth.

We use reciprocal space maps (RSM) to get access to the in-plane lattice
parameters of the sample. In order to check both the \textit{a} and
\textit{b} lattice parameters, we map the off-specular (1 0 7) and
(0 1 7) H and K directions of the RS respectively. The results for
a 4 unit cell thick film grown on SLAO are exhibited in Fig. 3, which
shows the out-plane (L) direction of the RS plotted vs. the in-plane
one. (Fig. 3A corresponds to the K direction and Fig. 3B corresponds
to the H direction.) The intense peak in both maps corresponds to
the SLAO substrate reflection while the thin film reflection is the
wide (narrow) maximum along the Y axis (X axis). The maps confirm
that the sample is epitaxial and are evidence of an in-plane coherence
length of the thin film limited by the substrate in-plane coherence
length. The wide thin film peak in the L direction is a consequence
of the finite size thickness (\textasciitilde{} 5.5 nm) of the sample. 

\begin{figure}
\includegraphics[width=7.5cm]{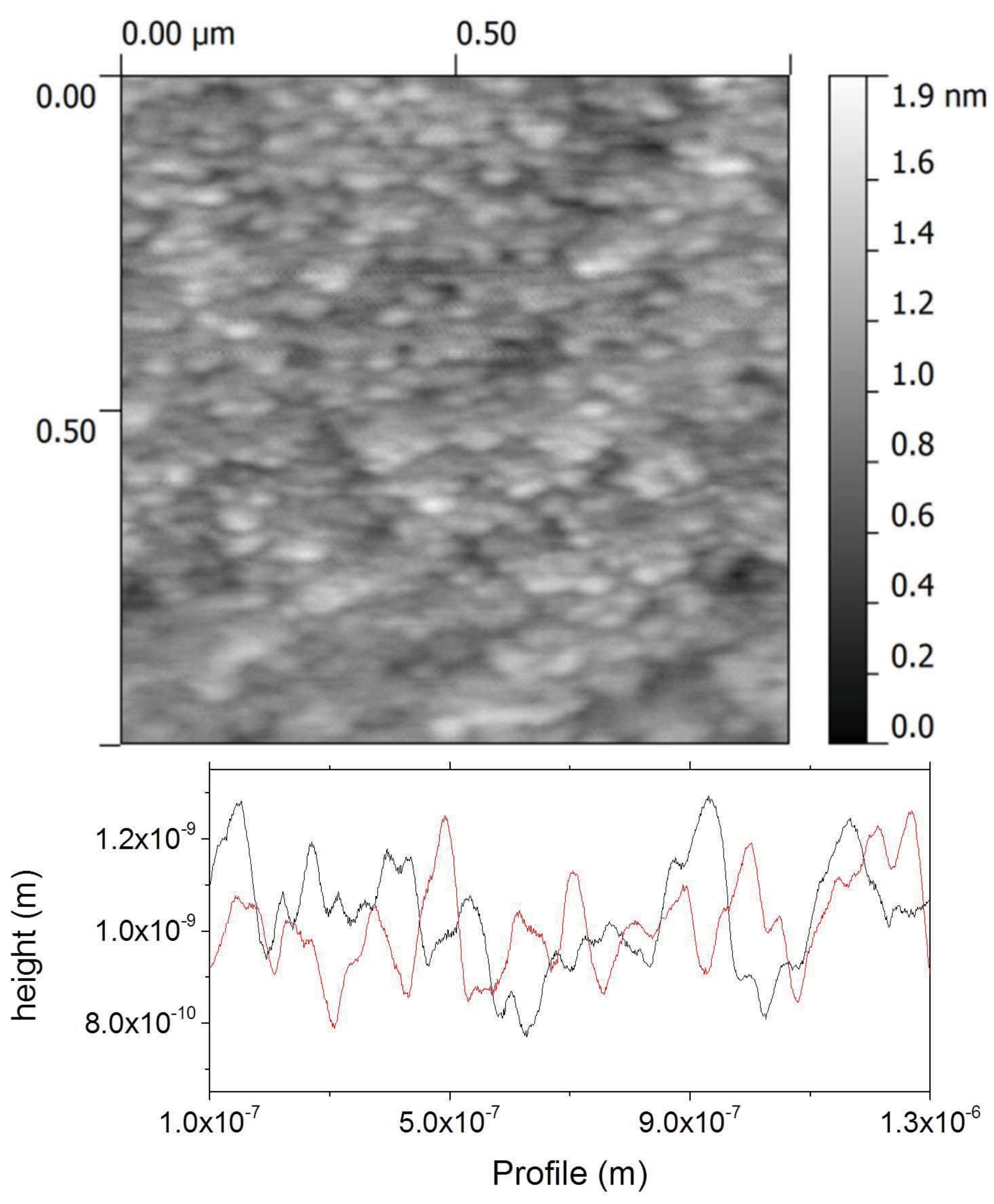}\caption{(Color online) (Top image) Surface height AFM scan of the 4 unit cell
thick film (5.3 nm). (Bottom graph) Cross-sectional analysis of the
diagonal cuts of the top image}
\end{figure}

Figure 4A shows a 1 $\mu$m $\times$ 1 $\mu$m AFM scan of the 4
unit cell fi{}lm. The average root mean square roughness, $\sigma$$_{rms}$
is 0.2 nm, which is strictly the same as that obtained for the 10
$\mu$m $\times$ 10 $\mu$m scans. The thickness variation relative
to the total thickness of the sample is 3.75 \%. In Fig. 4B the height
profile scans obtained from the main diagonals of the image can be
observed. They reveal that the largest height variation is $\sim$
4 Å, which corresponds to a third of a $\delta$-LCO unit cell. The
surface roughness of the thin film matches the surface roughness of
the SLAO substrate after it is annealed under conditions equivalent
to those of the growth.

\section{Electronic Double Layer Transistor fabrication and operation}

All the set of transport experiments described in this work were carried
out using the same EDLT. The device was fabricated using a 4 unit
cell ($\sim$5 nm) thick film of $\delta$-LCO. To preserve the sample
surface during the fabrication of the device, the cuprate film was
kept in an inert N$_{2}$ gas atmosphere inside a small portable vacuum
chamber. When procedures open to the atmosphere were required, a continuous
flow of the same gas was used to minimize the exposure of the surface
to moisture, CO$_{2}$ and other atmospheric gases that would potentially
degrade the surface quality of the thin film. Silver electrodes were
deposited on the four corners of the thin film to allow us characterize
its electrical transport properties. The electrodes were deposited
by a thermal evaporation method and during this process the sample
was kept at liquid N$_{2}$ temperature to avoid the de-oxygenation
of the thin film. As sketched Fig. 5a, the sample with the deposited
electrodes was mounted in a glass cylinder that contains the DEME-TFSI
ionic liquid used as the gate dielectric. Electrical wires were attached
to the Ag electrodes with indium dots and a Pt coil (gate electrode)
was installed on the top of the device. 

\begin{figure}
\includegraphics[width=8.6cm]{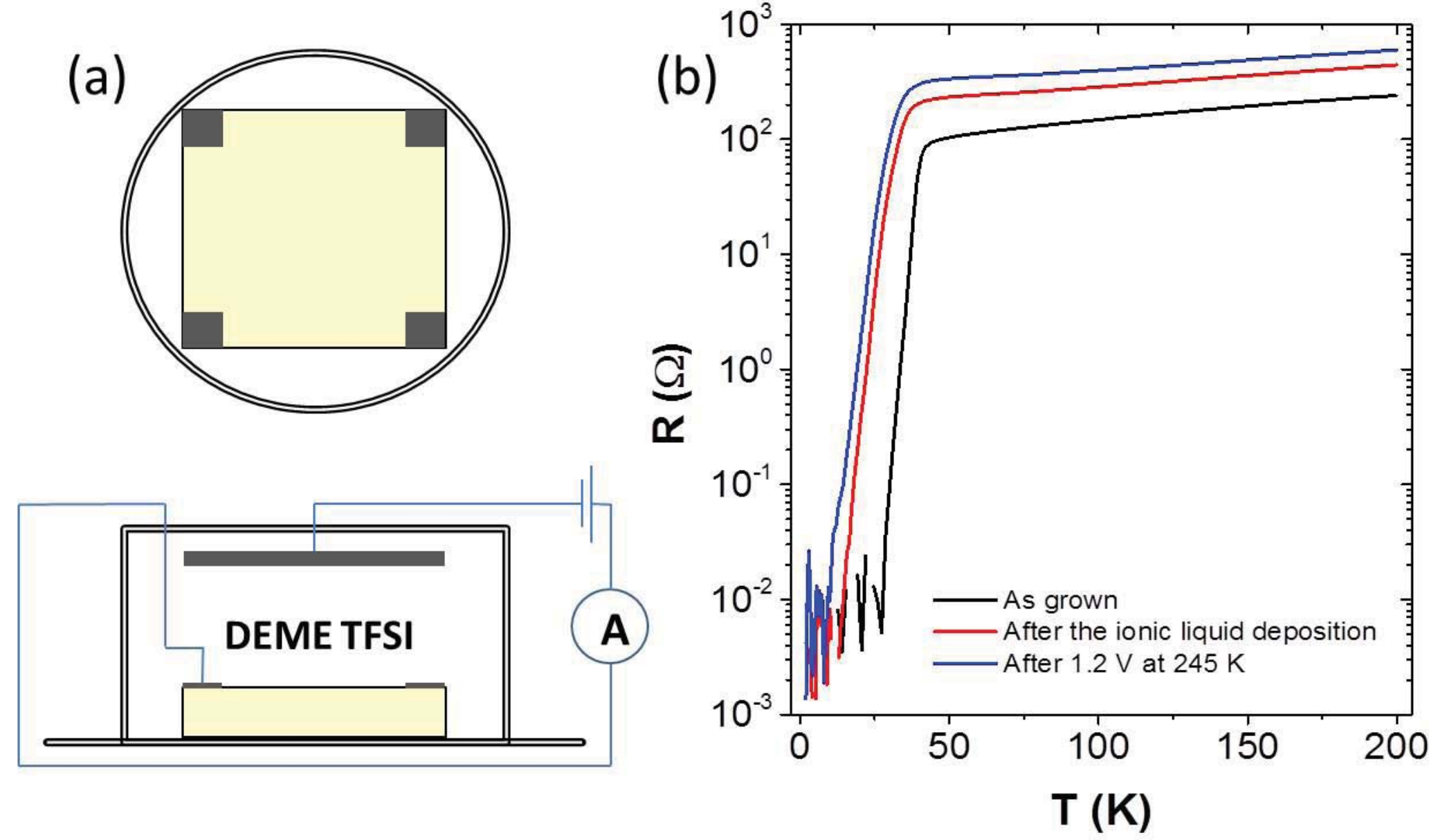}\caption{(Color online) (a) Sketch of the EDLT device. The top and bottom pictures
are the top and side views of the device respectively. (b) Resistance
versus temperature curves of the LCO thin film during the device fabrication
and optimization process.}
\end{figure}

In Fig. 5b we show the evolution of the R(T) curves of LCO film during
the fabrication and optimization of the device. Before applying the
ionic liquid to the surface of the film, and after the above described
manipulation, the as-grown sample has a R(T) curve characteristic
of a high quality $\delta$-LCO crystal. The superconducting transition
temperature (T$_{C}$) of the sample is $\sim$45 K which is the highest
T$_{C}$ reported for thin films of this thickness\cite{Bozovic2002}
or for the bulk form of this material.\cite{Chou1992} Once the ionic
liquid is added to the cylinder glass container we again measured
R(T) of the film (red curve). There is a drop in the onset temperature
of the superconducting transition of about 10 K and the transition
becomes broader. The ionic liquid is added at room temperature and
we cannot avoid what appears to be a slow room temperature chemical
degradation of the sample surface, which is completely arrested when
the sample is cooled down to the gating temperatures (240-245 K).
We start electrostatically charging the $\delta$-LCO film at 240
K by applying positive gate voltages (V$_{G}$) that induce negative
charges in the film, and therefore, reduce its carrier concentration.
The gating process at this temperature and up to 1.2 V does not produce
any significant change in the normal state resistance and does not
affect T$_{C}$ . In order to build up more charge on the sample surface
and enhanced the charge transfer to the sample, we increased the ionic
mobility by increasing the gating temperature to 245 K. After recovering
the previous V$_{G}$, i.e. 1.2 V, the normal state resistance of
the sample had increased by a factor of 1.5, and T$_{C}$ dropped
more than 1 K (the blue line in Fig. 5B). Subsequently, all the different
values of V$_{G}$ or hole concentration presented in this work were
obtained after gating the sample at 245 K over a period of 10-15 min.

In order to monitor the electrostatic charging of the sample, we measured
the charge leakage current between the Pt coil and the sample during
the gating process with the configuration illustrated in the bottom
panel of Fig. 6a. A Keithley 6517 electrometer was used as both the
in-series voltage source and as a picoammeter. An example of the leakage
current vs. time measurements is shown in Fig. 6a. After removing
the constant background which appears for the longer times of the
measurement, we integrated the curves as a function of time to estimate
the amount of charge transferred to the sample in each charging step
(see the inset of Fig. 6a). This allows us to calculate the total
accumulated charge for each increment of gate voltage. These are the
red symbols in Fig. 6b. Finally, the concentration of holes at the
$\delta$-LCO thin film for each V$_{G}$ (black symbols of Fig. 6b)
can be derived by assuming that the initial concentration of holes
per superconducting oxygen plane (holes/Cu) is given by the Hall number
of the initial V$_{G}$ measured at 190 K. The initial density of
holes for the 1.2 V gated thin film at 245 K is 0.1215 holes/Cu and
this is in good agreement with the estimation of the number of holes
obtained from the parabolic relationship between T$_{C}$ and the
hole concentration for the same conditions.\cite{Tallon1995} However,
we would like to clarify that this is not an accurate estimation of
the absolute value of charge density in the film, but it allows us
to estimate the relative evolution of carrier concentration with V$_{G}$. 

\begin{figure}
\includegraphics[width=7.5cm]{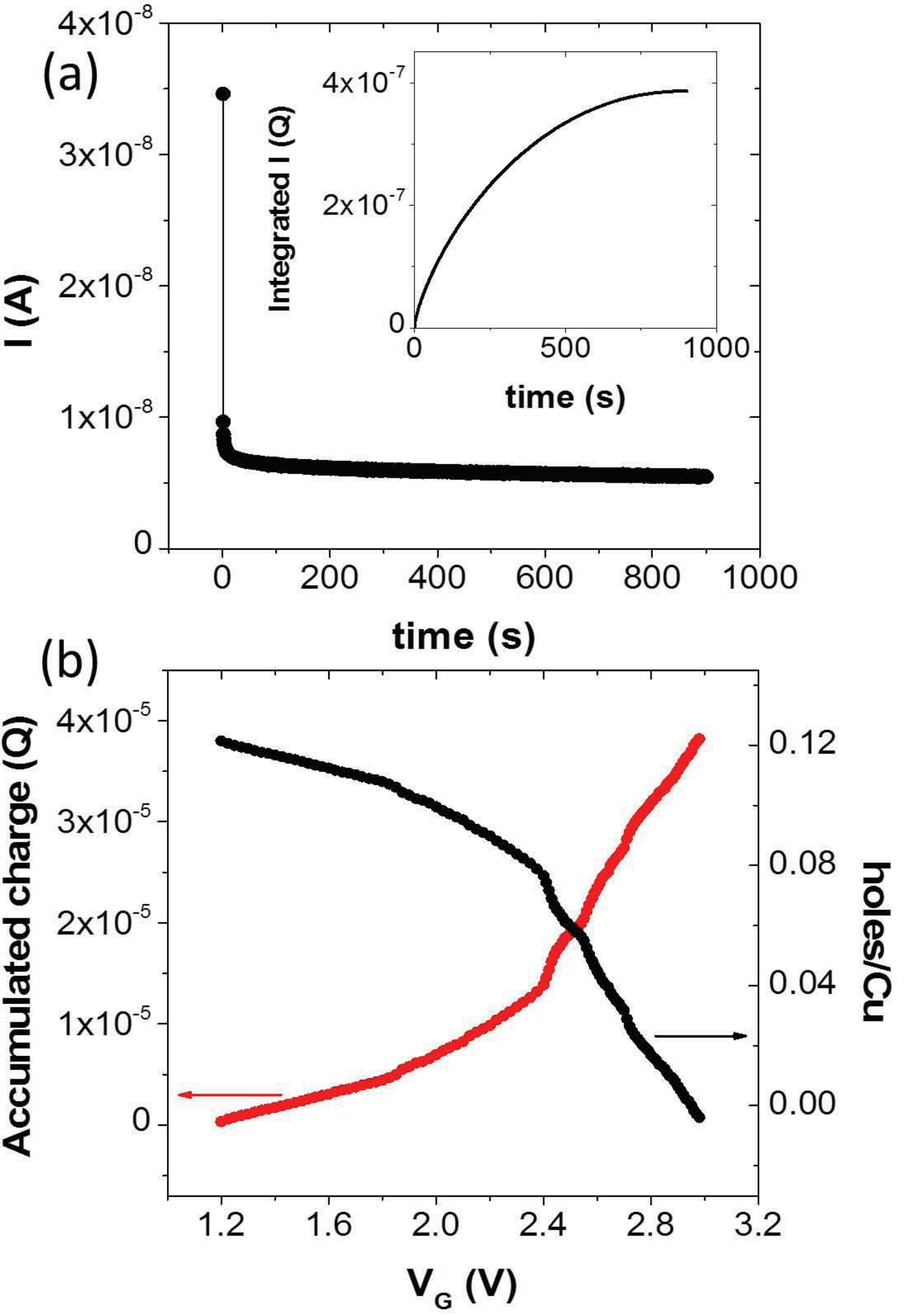}\caption{(Color online) (a) Leakage current measurement as a function of time.
The data was taking during the 2.15 V gating process. Inset: Integration
of the leakage current measurement after background subtraction. (b)
Accumulation of charge during the complete experiment (left y axis
and red dots) and total density of charge carriers per superconducting
Cu (right y axis and black dots).}
\end{figure}

\section{Transport and magneto-transport measurements. Results and Discussion}

The electrical resistance of the sample was measured using a four-terminal
configuration employing the following devices: a Keithley 7001 switching
system, a Keithley 6221 current source and a Keithley 2182A nanovoltmeter.
The temperature of the device and the magnetic field conditions of
the measurement were controlled in a Physical Properties Measurement
System (PPMS). 

We designed an experiment to access three different measurement configurations
of the in-plane resistance based on a four-terminal transport experiment.
The first two resistance configurations were obtained by applying
the electrical current in the main tetragonal in-plane crystallographic
directions of the film that coincide with the edges of the sample.
We denote these resistances as R1 and R2, as well as their sheet resistance
counterparts as Rs1 and Rs2 respectively. In order to determine the
values of sheet resistance along the main crystallographic directions
of the $\delta$-LCO thin film (Rs1 and Rs2) from R1 and R2, we applied
the method explained in the reference.\cite{DosSantos2011} The method
is a revision of the Montgomery method,\cite{Montgomery1971} and
it allows us to determine the 2D electrical sheet resistance tensor
of the tetragonal structure of the $\delta$-LCO thin film. The third
resistance was the Hall Resistance denoted by Rxy. We have characterized
the three resistances in zero field and in a 9 T applied magnetic
field. This measurement configuration allows us to obtain the ratio
Rs1/Rs2, which will be used to evaluate the in-plane anisotropy of
the electrical properties. 

\begin{figure*}[t]
\includegraphics[width=12cm]{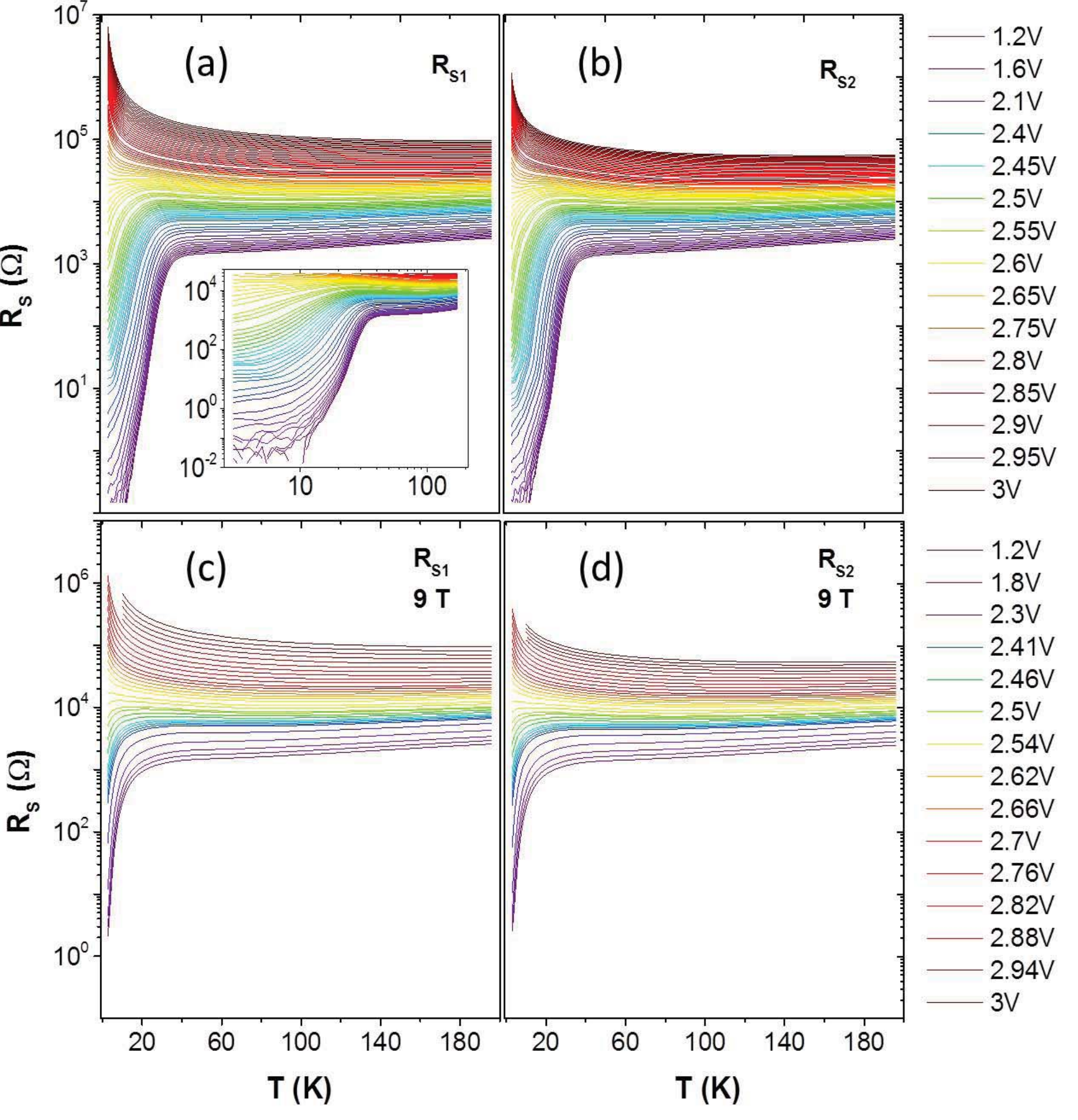}\caption{(Color online) Sheet resistance as a function of temperature for different
values of V$_{G}$ and in two different magnetic fields. Figs. (a)
and (c) show the data taken in the Rs1 direction with 0 T and 9 T
magnetic fields respectively, and Figs. (b) and (d) show the measurements
obtain for the Rs2 direction under 0 T and 9 T magnetic fields as
well. The inset of Fig. (a) shows a magnification of the superconducting
part in a log-log scale.}
\end{figure*}

Figure 7 shows a false color plot of the dependence on temperature
of Rs1 and Rs2 from 3 K to 200 K for different values of V$_{G}$
for both 0 T and 9 T magnetic fields. The different values of V$_{G}$
range from 1.2 V (purple or cold) to 3 V (brown or hot) corresponding
to a concentration of holes per copper oxygen plane of 0.1215 holes/Cu
and (-0.00625) holes/Cu respectively. Figures 7a and 7c show the data
obtained for Rs1 in zero field and in a 9 T magnetic field, respectively,
and Figs. 7b and 7d show the results obtained for Rs2 under the same
field conditions. We display a total of 75 different values of V$_{G}$,
i.e. doping levels, for the data obtained without magnetic field (for
both Rs1 and Rs2) and 33 values of V$_{G}$ for the sets of data taken
at 9 T. The legend displays the color code for some representative
values of V$_{G}$. 

As can be seen in Fig. 7, both curves of $Rs1(T)$ and $Rs2(T)$ show
a characteristic evolution from superconducting to insulating behavior
as V$_{G}$ is increased and the charge carrier concentration is depleted.
For V$_{G}$=1.2 V, the sample is metallic down to the superconducting
transition temperature T$_{C}$$\sim$40 K for the 0 T data and down
to T$_{C}$$\sim$25 K for the 9 T data. With a further increase of
V$_{G}$ the transition temperature slowly moves towards lower temperatures
as expected for the depletion of holes in the superconducting CuO$_{2}$
planes. Below 40 K (25 K) the transition is continuous and the corresponding
gate voltage that separates the superconducting and insulating regimes
is V$_{G}$=2.64 V (V$_{G}$=2.54 V) for the measurements obtained
with a 0 T (9 T) magnetic field. Under 0 T and 9 T conditions of applied
magnetic field, Rs1 and Rs2 show a similar evolution with V$_{G}$.

\subsection{Finite size scaling }

These data of resistance vs. temperature at different gate voltages
resemble that of the reported continuous SITs as a function of charge
carrier concentration in the cuprates YBa$_{2}$Cu$_{3}$O$_{7-\delta}$\cite{Leng2011}
and La$_{2-X}$Sr$_{X}$CuO$_{4}$.\cite{Bollinger2011} Continuous
SITs are quantum phase transitions (QPTs) with quantum critical points
separating the superconducting and insulating ground states. Near
the QCP and at nonzero temperature the different phases are separated
by a Quantum Critical Regime (QCR). In the QCR, the dependence of
a macroscopic observable property, Rs, as a function of the tuning
parameter, the concentration of holes per superconducting CuO$_{2}$
plane (p), should follow a scaling law of the form:

\begin{equation}
Rs(p,T)=Rs_{C}f(|p-p_{C}|T^{-\frac{1}{\nu z}})
\end{equation}

Here $Rs_{C}$ and $p_{C}$ are the critical sheet resistance and
the critical carrier concentration respectively, T is the temperature
and $\nu z$ is the critical exponent product that characterizes the
universality class of the phase transition.\cite{Sondhi1997}

Following the procedures described in the reference,\cite{Markovic1999}
we first obtained the critical values of carrier concentration for
the four different configurations of the experiment. After plotting
Rs as a function of p for the lower temperature isotherms of the experiment
(3-10 K), we found a crossover, which is the value of the temperature
independent sheet resistance at critical doping, $p_{C}$. (See the
inset of Fig. 8a as an example.) The value of $p_{C}$ for the 0 T
measurements is 0.04 holes/Cu and for the 9 T measurements is 0.055
holes/Cu. It is independent of the in-plane direction of the measurements.
We then obtain $\nu z$ by the evaluation of the derivative of Rs
with respect to p at $p_{C}$. This procedure involves the fitting
of all the isotherms of the study by a polynomial function that accurately
takes into account the curvature of the data at this point. The results
of the polynomial fit are the lines plotted in the inset of Fig. 8a.
We then obtain the critical exponent product of the transition as
the negative of the inverse of the slope of the straight line yielded
from the log-log plot of $(\frac{\partial Rs}{\partial p})_{p_{C}}$
vs. $T$. This is shown in Fig. 8a along with the linear fits to the
data. Interestingly we obtained two different values of $\nu z$ for
the 0 T measurements: $\nu z$ =1.2 for Rs1 and $\nu z$ =1.4 for
Rs2, while the transition under a 9 T magnetic field is characterized
by a single value of $\nu z$ for both directions of the applied electric
field: $\nu z$ = 1.8. As can be observed in Fig. 8a, this method
is a sensitive tool for determining $\nu z$ , allowing us to distinguish
between small variations of the slope. Moreover, it allows us to estimate
the upper temperature limit of the QCR as the temperature at which
the analysis deviates from the linear regime. Surprisingly, the quantum
fluctuations persist up to approximately 30 K.

\begin{figure}[h]
\includegraphics[width=7.5cm]{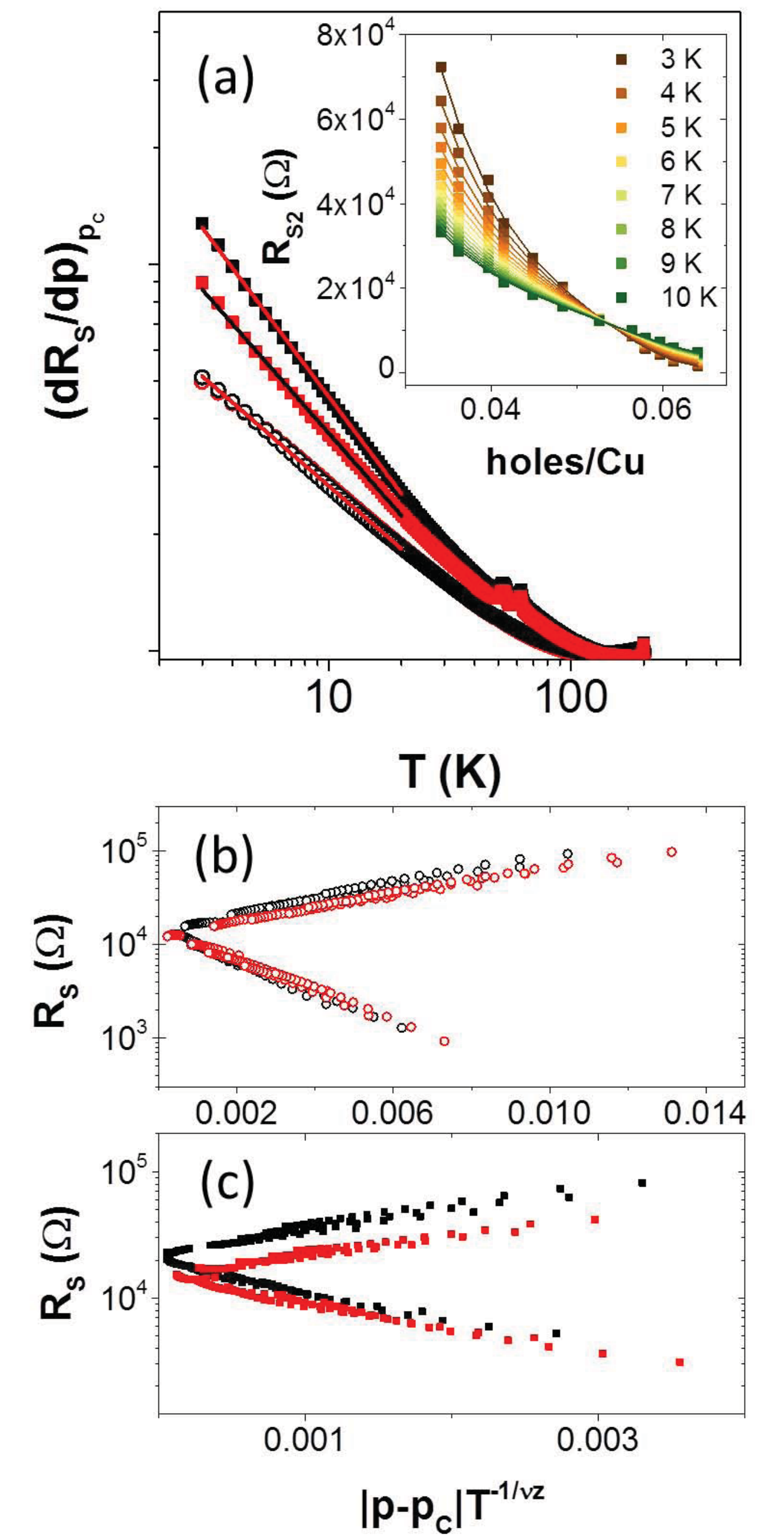}\caption{(Color online) (a) Value of the normalized derivative of RS as a function
of p evaluated at the critical concentration of charge density for
Rs1 (red) and Rs2 (black) in both 0 T (solid squares) and 9 T (open
circles) magnetic fields. Inset: Sheet resistance as a function of
charge density for 15 isotherms ranging from 3 to 10 K (selected temperatures
are indicated in the legend). In the left panel the sheet resistance
as a function of the scaling variable is shown for data obtained without
magnetic field (b) and under a 9 T field. (c). The scaling is shown
for 6 (12) different hole concentrations ranging from 0.046-0.034
(0.064-0.034) for data obtained with 0 (9) T. The temperatures for
both calculations range from 3 to 10 K.}
\end{figure}

From the previous analysis of the data we obtain the values of Rs$_{C}$.
The critical value for Rs1 / 0 T (Rs2 / 0 T) is approximately 21.5
k$\Omega$ (15 k$\Omega$), and for Rs1 / 9 T (Rs2 / 9 T), RsC decreases
to 13.5 k$\Omega$ (12.5 k$\Omega$). The measured values of Rs$_{C}$
do not agree with the quantum resistance for pairs (6.45 k$\Omega$).
The result supports the possibility of a QPT driven anisotropically
in the absence of magnetic field and isotropically in its presence. 

An alternative method for determining the critical exponent product
of the transition is to obtain the best collapse of the Rs data inside
the QCR onto the scaling function.\cite{Markovic1999} In Figs. 8b
and 8c we show the results of this analysis for the 0 T and 9 T transitions
respectively. It highlights the collapse of the data as well as the
agreement of the values of the critical exponents obtained by the
two independent methods. This analysis suggests that there is a direct
SIT between superconducting and insulating regimes with the caveat
that the analysis is cut off at a relatively high temperature leaving
open the possibility of a more complex scenario at temperatures below
those actually measured. Indeed, a closer inspection of the $Rs(T)$
curves reveals that zero resistance is only achieve for the few highest
doping levels (see inset of Fig. 8a). Otherwise, the low temperature
resistance of the sample saturates reaching a constant value different
from the expected 0 $\Omega$ of the superconducting state. This result
could be a consequence of the occurrence of quantum fluctuations that
would prevent the system from becoming superconducting above 0 K and
that would be enhanced by the 2D character of the sample.\cite{Yen-Hsiang1986}

The SIT in 2D has been the focus of many experimental and theoretical
works for several years. These include studies of amorphous Bi,\cite{Parendo2005}
amorphous MoGe,\cite{Yazdani2000} InO$_{X}$,\cite{Sambandamurthy2004}
TiN,\cite{Baturina2007} STO,\cite{Caviglia2008} and more recently
in LSCO,\cite{Bollinger2011} YBCO\cite{Leng2011} and FeSe.\cite{Schneider2012}
To our knowledge there has not been any previous report of a possible
anisotropic QPT that, moreover, becomes isotropic under an applied
magnetic field. This result would imply that there is an intrinsic
electronic anisotropy in the sample that drives the quantum phase
transition with a different mechanism depending on the in-plane direction.
A magnetic field perpendicular to the CuO$_{2}$ planes is able to
suppress the anisotropy suggesting a link between the superconducting
state and the anisotropy. To further explore this phenomenon we characterized
the in-plane anisotropic electronic properties of the sample by means
of the coefficient Rs1/Rs2.

\subsection{In plane anisotropy of the sheet resistance }

Figures 9a and 9b show false color contour plots of the ratio Rs1/Rs2
as a function of carrier concentration and temperature for the data
obtained in zero field and in 9 T. In both maps, cold colors (violet)
represent a value of Rs1/Rs2=1, with the value of Rs1/Rs2 increasing
with the increase of the color temperature up to a maximum value Rs1/Rs2=5.4
(red.) It is shown that the electrical resistance of the sample in
the absence of magnetic field is isotropic in-plane for a wide range
of temperatures and hole carrier concentrations. For the highest level
of doping, p$\sim$0.12 holes/Cu, anisotropy appears once the sample
is cooled down to T$_{C}$. As the charge density decreases, the intensity
of the anisotropy increases and shifts towards lower temperatures
following the line of the superconducting dome. At the high carrier
concentration and low temperature portion of the color map, Rs1/Rs2
reaches a maximum value of 4 at p=0.06 holes/Cu and T=3 K. With further
depletion of holes, the intensity of the anisotropy falls and it achieves
a minimum at the critical carrier concentration of holes for the SIT
and at the lowest temperature of the experiment, i. e., p=0.04 holes/Cu
and T= 3K. From this point downward to the lowest carrier concentrations
the anisotropy continuously increases up to the highest value Rs1/Rs2=5.5,
deep inside the insulating regime. In this region of doping the anisotropy
is present up to 200 K and it decreases as the temperature increases. 

After applying a 9 T magnetic field perpendicular to the CuO$_{2}$
planes (Fig. 9b), the anisotropy of the high charge density part of
the color map is shifted downward, disappearing from the temperature
window of the experiment, in a similar manner to the suppression of
T$_{C}$. Interestingly, the rest of the contour plot remains strictly
the same as the one obtain for 0 T and the only differences emerge
after cooling the sample below 40 K (The coefficient of both Rs1/Rs2
obtained with 0 and 9 T is shown in Fig. 9c.)

The presence of in-plane anisotropy in the transport properties is
a striking result. The anisotropy is developed at the superconducting
transition which probably is enhancing the sensitivity of its detection.
Moreover, the presence of anisotropy would imply that the macroscopic
symmetry of the sample must be broken in order to set a preferential
channel for conductance along one of the Cu-O-Cu directions. In this
regard, the structural characterization of the sample has not revealed
any structural or morphological difference along the a and b directions
of the thin film. Neither AFM nor X-Ray experiments suggests this
possibility. However, we have no tool to characterize the oxygen interstitial
order in the ultra-thin film and we may be missing an important piece
of information.

\begin{figure*}[t]
\includegraphics[width=12cm]{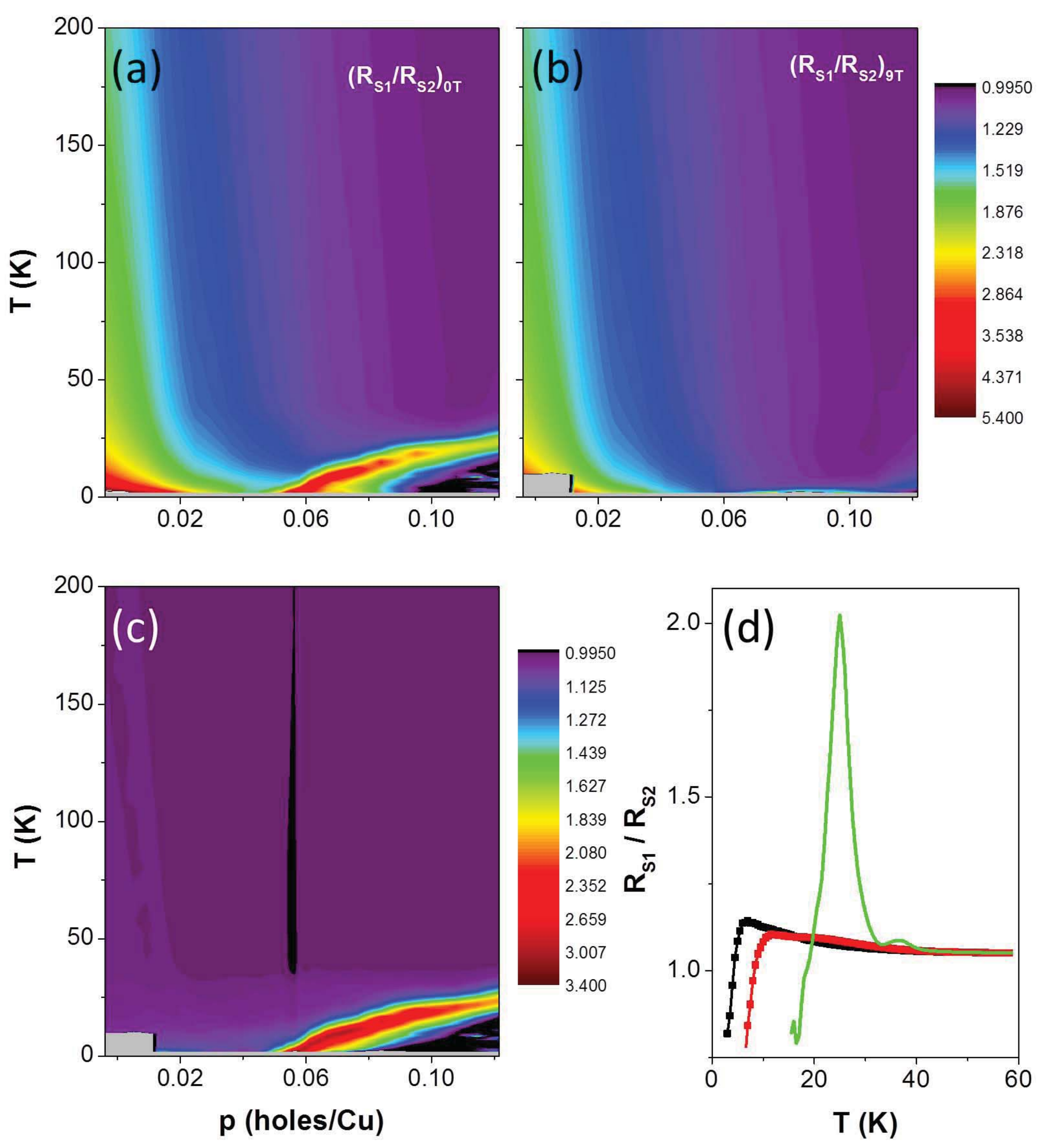}\caption{(Color online) False color contour plots of the coefficient of Rs1/Rs2
obtained with 0 T (a) and 9 T (b). The purple color represents a value
of Rs1/Rs2\textasciitilde{}1 and the red color represent a value of
Rs1/Rs2\textasciitilde{}5. (c) False color contour plot of the coefficient
of the 0 T and 9 T Rs1/Rs2 values. (d) Rs1/Rs2 value vs. temperature
on different magnetic fields: 0 T (green line), -4.5 T (red squares),
-9 T (black squares), 4.5 T (red line) and 9 T (black line) for a
charge concentration of 0.1215 holes/Cu.}
\end{figure*}

The observation of a quantum SIT that behaves differently depending
on the in-plane direction, together with the emergence of an in-plane
anisotropy in the electrical properties of the sample, is evidence
that the electronic ground states of $\delta$-LCO thin films are
intrinsically anisotropic on both sides of the QCP. Inside the QCR,
the quantum fluctuations between both regimes become more important.
They eventually result in a reduction of the electronic anisotropy
and the system correspondingly becomes more isotropic as it becomes
more quantum in nature (i.e. Rs1/Rs2 decreases as we approach the
QCP.)

One of the possible ways of interpreting the existence of a preferred
conductance channel in the system can be done in terms of the models
of electronic and spin orderings and the experimental observations
of stripe formation in the 214 family of cuprates.\cite{Berg2009}
In this scenario, the charge that dopes an antiferromagnetic Mott
insulator it is not randomly distributed in the sample. It forms ordered
clusters of charge and spins promoted by antiferromagnetic correlations
and charge density modulations. The interlocked patterns can vary
from a \textquotedblleft{}classical crystal (CC)\textquotedblright{}
order of charge on the insulating side, to \textquotedblleft{}quantum
electronic liquid crystal phases (QELC)\textquotedblright{} or charge
or spin density waves in the superconducting regime.\cite{KivelsonS.A.FradkinE.1998}
In this regard, neutron scattering experiments on stage 4 and 6 $\delta$-LCO
single crystals have shown that a spin density wave state appears
at the superconducting transition temperature, which is enhanced with
a perpendicular applied magnetic field, contrary to our observations
on electrical anisotropy in field.\cite{Khaykovich2003} In the system
under study, the 2D character of the ultrathin film could be enhancing
the essential 2D character of the superconducting CuO$_{2}$ planes,
which may result in a 1D chain clustering of holes. Accordingly, stripe
formation could produce a lower resistance channel for current flowing
along the chains of holes than for current flowing across them. 

The comparison of data with and without magnetic field is evidence
that the nature of the electronic ground states on the insulating
and superconducting sides are different depending upon the magnetic
field. On the one hand, at the lower levels of doping the anisotropy
is more stable and the 9 T field barely has any effect on its temperature
and charge density dependences. These results suggest that the lightly
doped insulating state could be antiferromagnetic, as might be expected
from the bulk behavior, and a long-range exchange interaction of spins
would stabilize robust charge and spin orderings, independent of magnetic
field. On the other hand, the effect of magnetic field on the superconducting
side reduces the intensity of the anisotropy significantly and it
only appears at the lowest temperatures. This result could be a consequence
of the reduction of T$_{C}$. As stated earlier, the superconducting
transition could be enhancing the sensitivity of detection and the
presence of the anisotropy in other regions of the phase diagram could
pass unnoticed. Nevertheless, the effect of magnetic field on the
anisotropy is evidenced in Fig. 9d. Here we have plotted the value
of Rs1/Rs2 for p=0.1215 holes/Cu measured with 0 T (green line), 4.5
T (red line), -4.5 T (red squares), 9T (black line) and -9 T (black
squares) as a function of temperature. The anisotropy is developed
only in the absence of field and is not correlated with the transition
temperature, which is shifted systematically to lower temperatures
with the increasing field. We suggest that the anisotropy could arise
due to electronic inhomogeneity and the possibility of strong spin-orbit
scattering in a highly 2D system could lead to the magnetic response
of the anisotropy. However, there is neither a specific model for
the anisotropy in the conductance of the thin film, nor a model for
its response to magnetic field.

\subsection{Hall Resistance }

In order to characterize the temperature and doping dependence of
Rxy of the LCO thin film, we measured the Hall voltage (V$_{H}$)
with a 9 T applied magnetic field. We did this for both positive and
negative currents across both of the diagonal contacts of the four-terminal
configuration when the current was applied in the opposite diagonal.
The same measurements were also carried out without magnetic field,
to check the zero value (see Fig. 10a), and with 9 T during the warming
of the sample from 3 K to 200 K. Accuracy tests under -9 T, -4.5 T
and 4.5 T were measured at selected carrier concentrations as a control
parameter of the experiment (see Fig.10b). This was also done in resistance
vs. magnetic field sweeps. The data confirm the linear dependence
of Rxy on magnetic field and that the sensitivity of the measurement
of 0.1 $\Omega$.

Figure 10b shows the temperature dependence of Rxy for different levels
of doping ranging from 0.12 holes/Cu (violet line) to 0.04 holes/Cu
(brown line). Over the complete range of carrier concentrations, Rxy
increases with decreasing temperature up to a maximum value. Upon
further cooling below the maximum region (T $\sim$ 15 K) and depending
on whether the sample becomes a superconductor or an insulator, the
value of Rxy keeps dropping or eventually increases. It is important
to note that the change of $Rxy(T)$ to a negative slope at the lowest
levels of doping and temperatures is correlated with non-zero values
of Rxy measured without magnetic field, and therefore is not meaningful.
We have further explored the previous result and obtained the maximum
position for each of the $Rxy(T)$ curves as well as for each isotherm.
These data are displayed in the inset of Fig. 10b and show that the
Hall resistance has a maximum at 0.04 holes/Cu at a temperature of
$\sim$70 K.

\begin{figure*}[t]
\includegraphics[width=12cm]{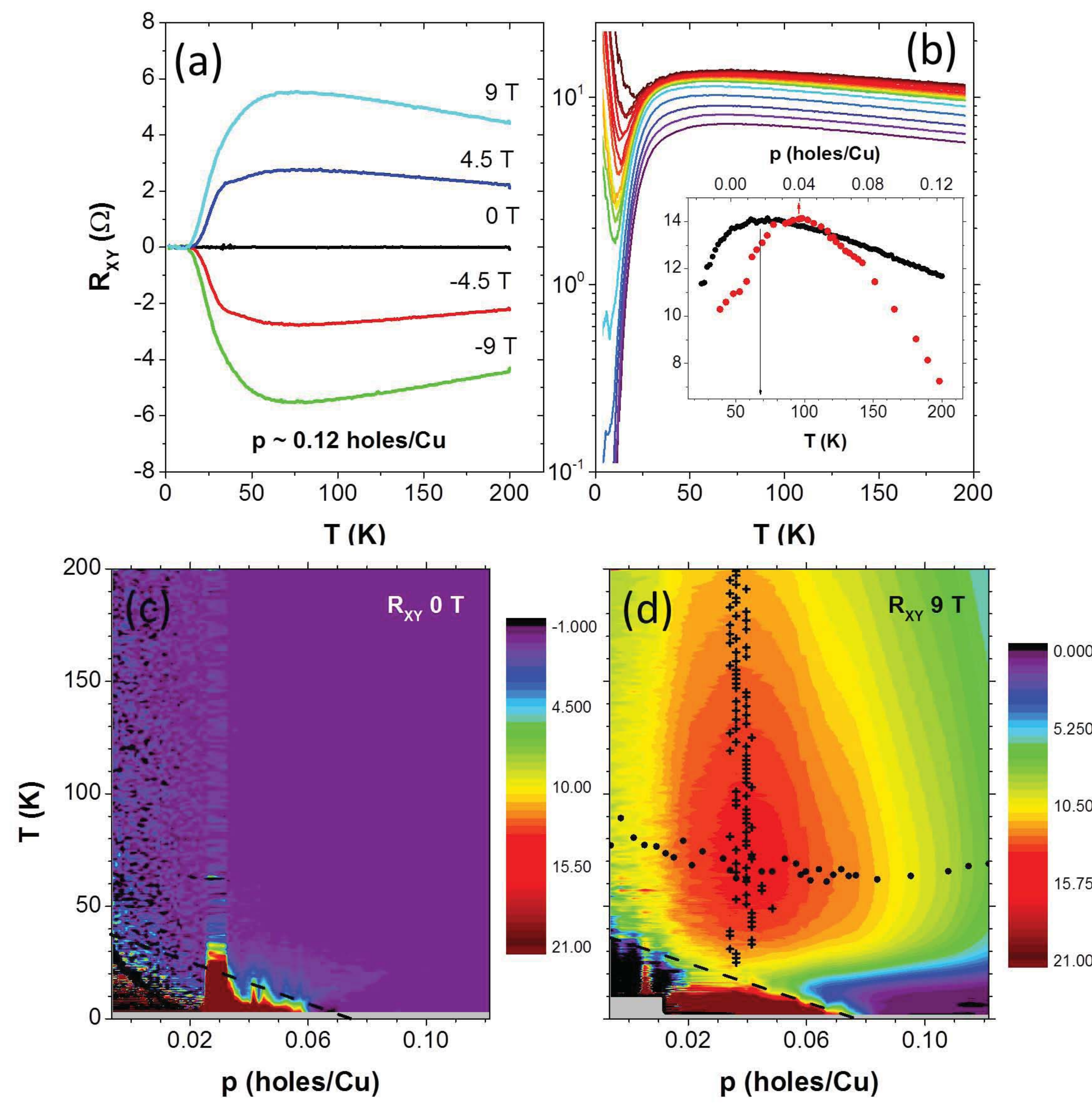}\caption{(Color online) (a) Hall resistance as function of temperature for
0 T (black line), -4.5 T (red line), -9 T (green line), 4.5 T (blue
line) and 9 T (cyan line). The measurements correspond to a charge
concentration of 0.1215 holes/Cu. (b) Rxy as a function of temperature
for different levels of doping ranging from 0.12 (purple) to 0.04
(brown) holes/Cu. The inset shows the dependences of the maximum of
Rxy as a function of temperature, for the different carrier concentrations
(black symbols, lower scale), and as a function of charge density,
for the different isotherms (red symbols, upper scale) are shown.
(c) and (d) Color contour plot of Rxy in the T-p plane. The data have
been obtained in a 0 T (c) and 9 T (d) magnetic field. Below the dashed
line the data is meaningful. The T and p dependences of the Rxy maximum
are plotted in (d) with crosses and dots respectively.}

\end{figure*}

To better visualize this surprising result we show a contour color
plot of the Rxy measurements in the temperature-charge concentration
plane (Fig. 10d). Note that the purple colors are assigned to Rxy$\sim$0
$\Omega$. As a result, Rxy measured with a 9T applied magnetic field
has a drop-like shape with the maximum values of Rxy represented in
red and located around the 0.04 holes/Cu-70 K point. Moreover, by
using this representation we can follow the evolution of the maximum
of Rxy in the T-p plane. The results reveal that the position of the
maximum as a function of p (the cross symbols in Fig. 10d) for the
different isotherms is roughly linear with a small shift to lower
p as the charge density decreases (from p=0.04 holes/Cu at the lowest
temperatures to p=0.035 holes/Cu at 200 K). When the maximum is evaluated
as a function of T for the different p values, it is a curve shape
with a minimum corresponding to p$\sim$0.08 holes/Cu at T$\sim$60
K.

The presence of a maximum in the Hall resistance could be revealing
an electronic phase transition,\cite{LeBoeuf2011}\cite{Leng2012}\cite{Balakirev2003}
which is a surprising and interesting result. The surprise comes from
the fact that the maximum value of Rxy is found at the critical carrier
concentration for the SIT at 0 T and the QPT would be a consequence
of an electronic transition which occurs even at high temperature.
At the temperatures where the maximum is observed, we have found that
the longitudinal resistances measured with 0 T and 9 T are strictly
the same. Hence, the Rxy maximum is evidence of a phenomenon that
promotes the QPT at 0 T. The interesting implication of this discovery
is that the SIT in electrostatically doped $\delta$-LCO would be
driven by electronic interactions since it would involve a change
in the electronic properties of the system when it transitions from
the insulating to the superconducting phases. This scenario is further
support by the presence of a minimum in the anisotropy of the electrical
transport properties at the QCP.

\section{Conclusions}

The results presented here highlight the role that electronic interactions
play in the transition from the superconducting state to the insulating
regime. The low temperature transport scales down to the lowest temperature
of the experiment, which suggests a possible QPT that is anisotropic
without magnetic field and isotropic in the presence of a 9 T field.
The electronic nature of this anisotropy is reveal by the measurements
of the coefficient Rs1/Rs2, which interestingly are minimized at the
critical concentration of charge carriers of the SIT. Furthermore,
at this concentration of holes we have found a maximum in the Hall
resistance at high temperature that could be revealing the electronic
nature of the low temperature transition. 
\begin{acknowledgments}
We thank the late Zlatko Tesanovic, Guichuan Yu, Wojciech Tabis, Shameek
Bose, Rafael Fernandes, Brian Skinner, Boris Skhlovskii, Steve Kivelson,
Vlad Dobrosavljevic, Marcello Rozenberg, Carlos Leon and Jacobo Santamaria
for helpful discussion; we also thank Chad Geppert for his help with
transport measurements. This work was supported by the National Science
Foundation under Grants No. NSF/DMR-0854752 and NSF/DMR-1209678. Part
of this work was carried out at the University of Minnesota Characterization
Facility, a member of the NSF-funded Materials Research Facilities
Network via the MRSEC program. J. G. B. thanks the financial support
through the Ramon y Cajal Program.
\end{acknowledgments}

\end{document}